# The Astronomical and Ethnological Components of the Cult of Bird-Man on Easter Island

Sergei Rjabchikov[1]

[1]The Sergei Rjabchikov Foundation - Research Centre for Studies of Ancient Civilisations and Cultures, Krasnodar, Russia, e-mail: srjabchikov@hotmail.com

**Abstract**[2]

The bird-man cult remains the main secret of Easter Island (Rapa Nui), a remote plot of land in the Pacific. This paper includes not only necessary ethnological data, but also some results on the archaeoastronomy. The research of some lines marked on a stone calendar from the Mataveri area, an archaic zone of the bird-man cult, allows to insist that the natives watched at least the stars Canopus and Aldebaran. There are strong grounds for believing that, among others, the Sun, the Moon as well as β and α Centauri were the matter for quasi-scientific enquiry. Several astronomical and calendar records in the rock art and in the script have been decoded.

**Keywords**: archaeoastronomy, bird-man cult, rock art, writing, Polynesian

## Introduction

Mulloy (1961, 1973, 1975) was the first who proved that different Rapanui ceremonial platforms and statues were oriented in some directions toward the sun.

Several articles of mine also are devoted to the archaeoastronomical investigations of the antiquities of Easter Island (Rjabchikov 1997a, 1997b, 1998a, 1998b, 1999, 2001, 2010a). I have attempted to understand some important features of the Rapanui bird-man cult in a number of works; some valuable results have been achieved (Rjabchikov 1996a, 1996b, 1997c, 2009a, 2010a, 2012a). In this work I continue studying the Rapanui rock drawings. The development of the methodology of the research in the domain of the Polynesian art is the mainstream of several articles of mine (Rjabchikov 1996c; 1997a; 1997d; 1998a; 2000; 2001; 2010b; 2011a; 2012b; forthcoming).

I use the nomenclature and tracings of the Rapanui classical inscriptions offered by Barthel (1958). In addition to that, a glyph was copied from the bird-man wooden figurine housed in the Peter the Great Museum of Anthropology and Ethnography (Kunstkammer), when I participated in the conference "Macklay Readings" in St. Petersburg in 1994. The studies are based on my own classification and translation scheme in deciphering the *rongorongo* signs (Rjabchikov 1987: 362-363, figure 1; 2010c: 22). Moreover, I always take into account the vocabularies and rules of alternating sounds of the Polynesian languages (cf. Tregear 1891: XIV-XXIV).

## The Local Sources Tell

The elections of the bird-man (*tangata-manu*) occurred on the island in the old times in the austral spring annually when sooty terns (*manu-tara*) were nesting on the Motu Nui islet. The victorious warriors (*mata-toa*) lived at Mataveri and waited for the birds. The human sacrifices were performed in the Ana Kai Tangata cave located nearby. Another centre of the feast was the ceremonial village of Orongo. In

---

[2] An earlier version of this paper, "The Cult of the Bird-Man on Easter Island: Religious, Ideological and Historical Implications", was read to a session of the 19th Annual Conference of the New Zealand Studies Association together with the Centre of Pacific and Asian Studies, Radboud University, Nijmegen, the Netherlands on June 28, 2013.

September the priests conducted different rites here. There they read the special *rongorongo* records on tablets. Certain incantations were dedicated to the deities *Haua* and *Makemake*.

Each warrior had one or more servants called *hopu*. These men swam to the Motu Nui islet and searched for eggs. A warrior who got the first egg at Orongo was declared there as a new bird-man. In the thoughts of natives this egg was the incarnation of the god *Makemake*. The head of the victor was shaved and painted red. The bird-man's arm that touched the egg was decorated with a strand of a red tape and a piece of sandalwood. At first the bird-man walked from Orongo to Mataveri, then he settled as a sacred person in a special house at the foot of the Rano-Raraku volcano or at the royal area Anakena (Routledge 1998: 258-265; Métraux 1940: 333-341; 1957: 130).

**What do the names *Makemake* and *Haua* mean?**

In compliance with Métraux (1940: 314), the famous *Makemake* was the Polynesian god *Tane* or *Tiki* in fact. According to Maori myths (Tregear 1891: 233), the brothers *Maui* were sons of a male called *Makeatutara*. Moreover, the youngest of them with the name *Maui-tikitiki* was the sun deity (Best 1955: 22). The Maori name *Makeatutara* contains the component *Make*, and the Rapanui form *Makemake* is the complete reduplication of that archaic name. Both names designate the solar deity and signify 'too bright, clear; look, glance,' cf. Maori *ma* 'white, clean,' Samoan *'au'aumama* 'clean; good-looking,' Maori *ke* 'differently to what one expected' and Rapanui *hana ke!* 'how hot!'.

The strange Rapanui character *Haua* was permanent companion of the god *Makemake*. It is valid to say that *Haua* denoted the moon goddess *Hina* (Rjabchikov 1987: 365). It was almost full moon (Rjabchikov 1989: 124).

In the Maori beliefs, a pair of principal deities, *Tane-matua* '*Tane*-parent' and his wife *Hine hau* (or *ahu*) *one* 'Woman-created-from-earth,' existed (Whataharo; Smith 1913: 143). Here Maori *hau* signifies 'to exceed; excess.' This term is pertinent to the ideas of pregnancy, fertility and abundance. The Rapanui goddess *Haua* was the same as the Maori goddess *Hine hau one*.

**Astronomical Simulations as the Main Clue to the Mystery: Part 1**

In compliance with my research (Rjabchikov 2010a), the heliacal risings of the stars β and α of the constellation Centaurus were important stellar markers of the appearance of the *manu-tara* birds. These celestial bodies first appeared in the morning sky several days apart, for example, on September 2, A.D. 1775 and on September 11, A.D. 1775. It must be emphasised that α Centauri is the third brightest star in the sky. Here and below I have used the computer program RedShift Multimedia Astronomy (Maris Multimedia, San Rafael, USA) to look at the starry heavens above Easter Island.

In the records on the Great St. Petersburg tablet (Pv 3) both stars are called **28 8** *Nga Matua* or *Nga Vaka* 'The boats' (Rjabchikov 1993a: 6), cf. Rapanui *Nga Vaka* (α and β Centauri; literally 'Many Boats'), see figure 1.

1(Pv 3): 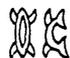

Figure 1.

Drawings of the *rei-miro* pendant (glyph **8** *matua*, in some cases *vaka*; cf. Maori *matua* 'hull of a canoe,' *motu tawhiti* 'ship') in some rock pictures designate the β and α Centauri; for example, in the Orongo rock picture decoded below (see figure 3) the *rei-miro* united with *Makemake*'s head denotes the heliacal risings of β and α Centauri.

In the folklore text "Apai" the month *Horahora* is mentioned: it describes the time when the sacral Canoe of *Tiki* and *Hina* arrived on Easter Island (Rjabchikov 1993b). It is the designation of the *Hora Nui* (September for the major part). The following sentence is interesting in the ancient text: *O(o) aku matua, o(o) aku matenga*. 'The ghost (*akuaku*) of the hull of the canoe enters, the ghost (*akuaku*) of the head (face) enters.' It means that first a star or several stars (the hull of the canoe) came, and then the sun (the head, the face/eyes) came. It is a poetical description of heliacal rising of this Canoe. Here Old Rapanui *matenga* means 'face; head,' cf. Rapanui *mata* 'face, eyes' and Maori *mātenga* 'head.'

The following inscriptions are written down on the Tahua tablet (Ab 3) and on the Santiago staff (I 6), see figure 2.

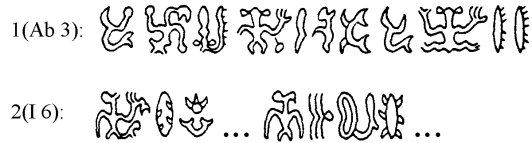

Figure 2.

1 (Ab 3): **2 62 73 50 6-15 50 19 8 2 69 46-46** *Hina too, hei Hora. Hiki Matua; Hina moko, naanaa.* 'The moon (the moon goddess *Hina*) took, (this goddess) drove the month *Hora* (*Hora iti* or *Hora nui*, August or September for the major part). The Canoe (β and α Centauri) was rising, and the moon (the moon goddess *Hina*) was invisible (hidden literally).'

In the Rapanui folklore the deities *Makemake* and *Haua* drove the *manu-tara* birds to the *Motu Nui* islet in September (Barthel 1957: 72). Old Rapanui *hei* means 'to drive,' cf. Rapanui *hei*, *hehei* 'ditto.' Old Rapanui *hiki* 'to rise, to lift' corresponds to Maori *hiki* 'to lift up; to raise.'

For example, during the heliacal rising of β Centauri (on September 2, A.D. 1775) the moon was invisible indeed, because it had already set.

2 (I 6): **44 (102) 54 82 7 8 … 44-33 47 (102) 28 …** *Takai Pipiri TUU Matua* [*Vaka*] *… tau avanga …* 'The STAR Canopus (and) the Canoe (β and α Centauri) are connected … (It is) the time of the grave (= the winter)...'

For instance, α and β Centauri as well as Canopus were visible in the sky on the night of June 15, A.D. 1812 (04:30).

The similar fragment of the "Apai" text reads as follows: *Ka Pipiri te hetuu tau avanga. Noi ruga Vake (= Vaka), noi runga Marua ua ...* '(It is) the star Canopus of the time of the grave (= the winter). (It is) the worship of the Canoe (β and α Centauri), (it is) the worship of (the first month) *Marua* (*Maro*; June chiefly) of rains...' (Rjabchikov 1993b; 2009a).

**An Additional Clue in the Art and Script**

One can try to realize the bird-man phenomenon studying the Rapanui rock art. Let us consider a panel at Orongo (Lee 1992: 102, fig. 4.97), see figure 3.

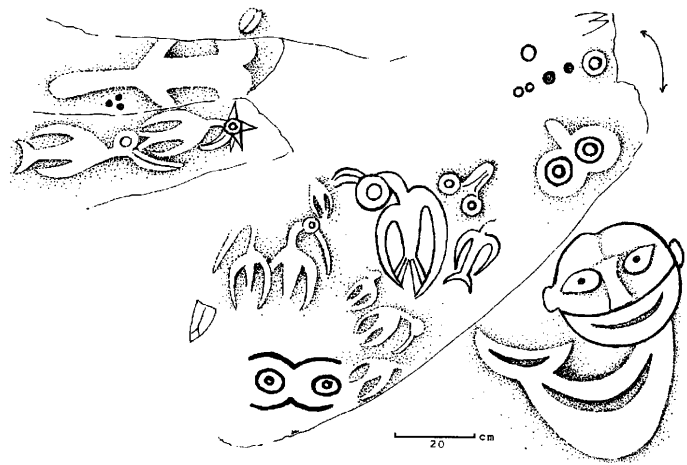

Figure 3.

Here a group of eight birds is represented. They all are shown with heads lowered, turned and even without heads. It is the symbol of sleeping and disappearance (glyph **44b** *tua*, cf. Rapanui *tua* 'back; behind,' Rarotongan *mokotua* 'back of a human being'). This terminology in a few place names may be linked to nights and nesting birds: *Tua te po* and *Tua a te manu* (Barthel 1962a: 107). Three large-eyed faces and one vulva surround these birds in the picture. These specific signs are glyph **1** *Tiki* denoting the power of the sun god *Tane* and glyph **60** *Mata* (Face), an image of the god *Makemake*, in other words, *Tane* or *Tiki* in the local *rongorongo* writing. The upper part of the picture on the left contains the following signs: a vulva, a lizard, three dots (cupules) near it as well as a bird and another bird connected with a star-like sign. So, two birds are represented near signs **1** *Tiki* (the sun deity), **69** *moko* (lizard; to hide) and **7** *tuu* (to come; star). In the Rapanui beliefs the Lizard (*Moko*) is an incarnation of the chthonic god *Hiro* (Barthel 1978: 251), and this word denotes the new moon in the local and Maori (New Zealand) calendars. The following motif is last in this composition: a pendant *rei-miro* depicting a boat is united with a head. This design is shown on the right.

The words of a Rapanui chant (Barthel 1962b: 854) illustrate the picture in general: *Ka memea no to Koro. Hami mea – tavake i tua e. Ka uuri no to Koro tangata – tuao i te Ohiro. Ka rava tangi no mahaki te makohe. Ka riti te hu pee o te kukuru toua, eve pepepepe. A ure Motu Nui. Etoru ange ra. Ka kai to Koro pera. Motu Nui, Motu Iti, Motu Kaokao.*

In my opinion, this text can be translated as follows: 'The red colour is from the Father (associated with) the dawn – the tropic bird is turned. The black colour (= the night) is from the Father – the bird *tuao* is on the day *Hiro* (= it is a description of a real or possible solar eclipse). The companion (= the moon goddess *Hina* or *Haua*) of the Frigate Bird (= the sun deity *Tiki* or *Makemake*) shouts aloud. An egg (*hu* = *hua*) of the Red Bird of the Eggs ([*Manu*] *Kura Toua*) which gives abundance becomes red, it is as a very fertile womb. It is a glyph **102** URE as a sign of fertility of the islet of Motu Nui. Here there are three directions (= they show three islets). The Father (= the sun) eats (the food) which is prohibited (for the people). These are islets of Motu Nui, Motu Iti, Motu Kaokao.'

Old Rapanui *hami* 'dawn; to dawn' corresponds to Rapanui *hamu* 'to dawn.' Old Rapanui *riti* 'red' corresponds to Rapanui *ritorito* 'red.' Old Rapanui *hu(a)*, *pu(a)* signify 'egg,' cf. Maori, Tahitian, Hawaiian *hua*, Samoan, Niue *fua*, Rarotongan *ua* 'egg' (< 'fruit'), cf. Rapanui *hua* and *pua* 'flower' as well. Old Rapanui *pe* means 'ripe,' cf. Rapanui *hakapee no kai* 'abundance of food,' and Tahitian *pe* 'ripe.' Old Rapanui *kukuru* 'red' corresponds to Rapanui *kura* 'red.' Old Rapanui *tohua*, *tohu* and *toua* signify 'egg,' cf. Rapanui *toua* 'yolk' and Tuamotuan *tooua* 'egg.' Old Rapanui *eve* 'womb' is comparable with Rapanui *eve* 'placenta,' Maori *ewe* 'placenta; womb; afterbirth,' Tuamotuan *eve* 'womb,' and Hawaiian *ewe* 'lineage; sprout.' Old Rapanui *ange* means 'direction,' cf. Samoan *ane* 'along.' The decoded archaic Rapanui terms are the key indicators here.

Let us examine two records inscribed on the New York wooden statuette of the bird-man, see figures 4 and 5.

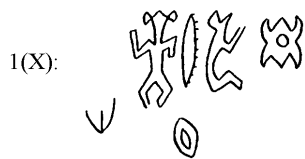

Figure 4.

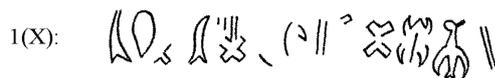

Figure 5.

The first text is written down on the belly, and it reads: 1 (X): **6-46 6-7 1 28** *Hana hatu Tiki, Nga*. '(It is) the heat of the lord *Tiki*, (it is) the egg.'

Old Rapanui *hana* 'heat; to heat; to shine' is comparable with Rapanui *hana* 'heat; to heat,' *mahana* 'day,' Maori *hana* 'to shine' and Tuamotuan *hana* 'the sun.' Old Rapanui *nga* 'egg's shell; egg' match Maori *nganga* 'shell, husk.'

*Tiki-te-hatu* (*Tiki* the lord) is mentioned in a Rapanui folklore text known as the Creation Chant (Métraux 1940: 321). This god of the sun, fertility and abundance was equal to the solar god *Tane* or was his procreative power. It is common knowledge that the people of the bird-man could burn the houses (Routledge 1998: 264). On the other hand, the god *Makemake* was related to the sun and to the fire in conformity with Ferdon (1961: 251).

The second text is written down on the bill, and this damaged text ends near the right eye. It reads: 1 (X): **5 25 49 5 49** [a segment is damaged] **49 4** [a segment is damaged] **41 7 44-4** *Atua hua. Mau atua, mau* [*atua*]*, mau atua ... retu, tatu*. '(It is) the lord (or deity) of an egg, the lord holds (it), the lord holds (it), the lord holds (it) … It is a tattooing (*tatu*, *retu*) on the head or the forehead.'

Here a warrior who obtained the first egg of the *manu tara* at Orongo and who after that was declared there as a bird-man is described. From that time on, he and only he held the egg on his palm.

One can suppose that each bird-man had a certain design on his face resembling glyph **25**. It must be underscored that in the design (Lavacheri 1939: plate IX, no 93; Mellén 1986: 190, photo 30) of the ceremonial stone at the site of Vai Tara Kai Ua glyph **25** *hua* (egg) is attached to the head of a bird-man. It is safe to assume that this glyph was the main tattoo mark of each bird-man.

The words of the same Rapanui chant (Barthel 1962b: 854-855) sound thus: *He Orongo no ta orongorongo. He retu no ta hu hatu retu.* In my opinion, this text can be translated as follows: 'The (place) Orongo (is associated with) the *rongorongo* (sacred readings). (It is) a tattooing (*retu*) looking like an egg (*hu* = *hua*) on the head or the forehead of the lord (*hatu*), it is such a tattooing.'

Let us examine the record inscribed on the St. Petersburg wooden statuette of the bird-man, see figure 6.

1(X2):    ||

Figure 6.

The text is written down on the neck, and it reads: 1 (X2): **4** *Atua*. '(It is) the deity (or the lord).'

**Bird-Men in Polynesia**

The first egg of the *manu-tara* birds from Motu Nui was an incarnation of *Makemake* according to Métraux (1957: 130). On the other hand, in the Maori religious system the god *Tiki-tohua* '*Tiki* – the egg' was an egg brought forth by *Hine-ahu-one*, creation and wife of the god *Tane* (Tregear 1891: 510). Hence, the god *Tiki* (representing as the vulva in the Easter Island rock pictures, cf. Rapanui *komari* 'vulva' < \**ko mari*) was equal to the holy egg (cf. Rapanui *mamari*, Old Rapanui *mari* 'egg'). It is a clue to the semantics of glyphs (petroglyphs) **1** *tiki*.

In a Tahitian song a warrior, handsome in death and became like a bird, was honoured (Alexander 1893: 58).

On Mangaia, the Cook Islands, and on Easter Island some men wore masks representing birds of the sun deity, maybe even frigate birds (Gill 1876: 49-50; Luomala 1977: 139; Routledge 1998: 268; see the interpretation in Rjabchikov 2011b: 5). Moreover, it is known that the Rapanui King *Nga Ara* wore a bird mask (Routledge 1998: 268).

It is interesting to note that the famous voyage of King *Hotu Matua*'s crew to Easter Island lasted from September 2 to October 15 according to Barthel (1978: 160). Hence, these immigrants could be compared with the *manu-tara* birds and even called bird-men.

I think that the Rapanui term *Nga Vaka* (α and β Centauri) was called so after the voyage of King *Hotu Matua*. Hence, the kings of the Miru tribe were the bird-men for long, and after the wars between the western and eastern tribes the competitions were invented to elect this holy ruler. King *Hotu Matua* had some features of the sun deity (Rjabchikov 1995; 2009b). Bird-men became the incarnations of the sun and fiery deity (*Makemake*, *Tane*, *Tiki*).

**On Some Names of Bird-Men**

In the *rongorongo* inscriptions there are at least two names of bird-men. In this connection, let us study the following record on the Great St. Petersburg tablet (Pr 6), see figure 7.

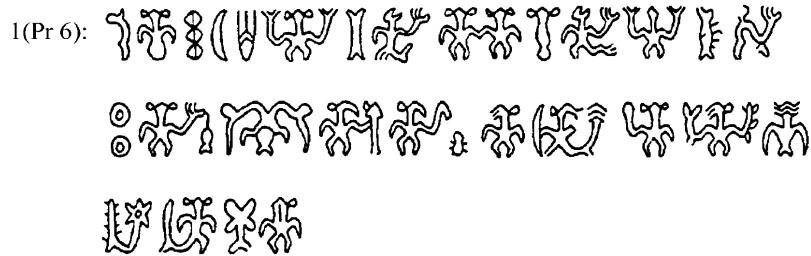

Figure 7.

The text reads as follows: **105 68 17 3 1 6-4 19 (102) 6-6 56 6-6 77 62 15 115-115 6 12 44-44 49 21 49 35 25 6 3 49 4/33 6 (123) 6-51 44-33 48 7 102 6 11 6 26** *Moe, hono te Hina, Tiki, hatu. Kio haha, paoa haha. Mama too ro takataka. (H)a ika tahataha mau oko, mau pa hua (h)a Hina. Mau atua/ua (h)a ake. Tau Utu ure a Mango-Ama.* 'The moon goddess (and the sun god) *Tiki* slept (and) were united, they produced (the eggs). The servant (*mata-kio*) took (the egg), the warrior (*paoa*, *mata-toa*) took (the egg). (The deity by the name of) *Maamaa* ('The bright colour') took the red hot colour. The bird-man ('corpse-frigate bird') held the egg ('ripe fruit'), (he) held the egg (*hua*) laying (on a piece of tapa) of the moon goddess. The deity (= the bird-man) held (a symbol) of abundance. (It was) the year of *Utu*, son of *Mango Ama*.'

This bird-man was called *Utu-Piro*; he ruled in ca. A.D. 1850 (Métraux 1940: 339).

I also have read the name of the bird-man *Ure te Ono* on the Berlin tablet (O 2[4]) and on Mazière's tablet (lines 3-4) (Rjabchikov 2012c). Both bird-men belonged to the royal Miru tribe. Routledge (1914-1915) has written numerous names of bird-men. Thus, one can say about the historical aspect of the bird-man cult.

**A Song about the *Manu-Tara* Eggs**

The next Rapanui chant (Campbell 1999: 217) has attracted my attention:

*Ka moe nga pua mo roto i te tama ere, mo hiki, mo turu ki te hongaa a pua. Teitei Renga o nga manu, Keu-Renga.*

The transliteration of the text and its interpretation are of mine: 'Many eggs (*pua*, *hua*) slept inside (the hiding places waiting for) the young man (= servant) for the elevating, for the coming down to the nests of the eggs. (The god) *Renga* (the yellow colour = the sun) of many birds grew, (it was) *Renga-Keu*.'

The key indicators here are Rapanui *hua* (*pua*) 'egg,' *hongaa* 'nest,' and *manu* 'bird.'

**Astronomical Simulations as the Main Clue: Part 2**

There are good grounds to believe that stars Aldebaran (α Tauri) and Canopus (α Carinae) were important celestial markers for the natives. It must be emphasised that the colour of Aldebaran is bright red. Besides, Canopus is the second brightest star in the sky. For the calculations I have chosen three years: A.D. 700, A.D. 1600 and A.D. 1850. The dates of the heliacal (first morning) rising of Canopus were May 17, A.D. 700, May 24, A.D. 1600 and May 25, A.D. 1850. The dates of the heliacal rising of Aldebaran were May 27, A.D. 700; June 12, A.D. 1600 and June 16, A.D. 1850. The new moons were on May 23, A.D.

700; on June 11, A.D. 1600 and on June 10, A.D. 1850. It is obvious that the heliacal appearance of Canopus predicted the beginning of the month *Maru* (*Maro*; the month of the winter solstice; rain season; it began in the new moon of June as a rule). The heliacal appearance of Aldebaran could be used to start the preparation for precise determination of the day of the winter solstice.

Mataveri was a well-known place where the warriors gathered before and after the main bird ceremonies at Orongo. In Mataveri they attended in different rituals in dancing (Routledge 1998: 259). There an observatory was situated. The local men themselves had to obtain better results. A number of lines were recognised on a huge rock by the runway of the airport in this area. Liller (1989) has measured the orientations of these lines, so that the azimuths of the setting sun for different days have been obtained. Among them the lines are which indicated the winter and summer solstices as well as the vernal equinox. Since the calculations are performed for the sun's positions, only three azimuths – 322.1°, 339.1° and 177.5° – are indeterminate.

Let us try to decipher them. For example, choose the year A.D. 1775. On December 20 (near the summer solstice) the azimuth of Aldebaran was 339.1° (23:44). On December 21 the azimuth of this star was 322.0° (00:44; the same night for the natives). The azimuth of Canopus was 177.5° (00:31) that night (Rjabchikov 2010a). It is obvious that the priests-astronomers looked at both stars during that and other nights.

Besides, with allowance made for the error (the corresponding azimuth was 286.7°), August 10 or August 11 was an important date in the Mataveri calendar. It is possible that the natives waited for the heliacal rising of the bright star Pollux (β Geminorum). It happened on August 10 from A.D. 1690 to A.D. 1720, and on August 11 from A.D. 1721 to A.D. 1796. I believe that on that day many warriors and priests met there and at Orongo.

With the same error (the corresponding azimuth was 277.9°), September 2 or September 3 was an important date in the Mataveri calendar. It might be the date of the heliacal rising of β Centauri, an important sign for waiting the *manu-tara* birds.

Consider two fragments on the Aruku-Kurenga tablet (Br 2-3, Br 3), see figure 8.

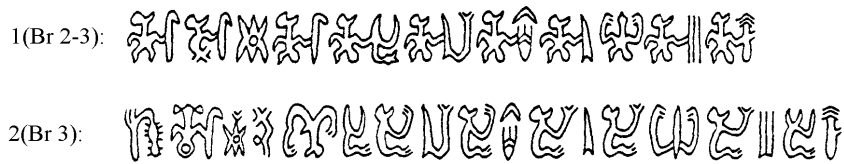

Figure 8

1 (Br 2-3): **6-35-6-35 7 6-35 6 2 6 5-15 6 1 3** (= a reversed variant) **6 5 6 53 6 34 6 9 33** … *Apaapa tuu, apa a Hina, a atua roa, a Tiki Hina uri, a atua, a Maru, a raa, a niva UA* …'The star (Aldebaran) appeared, the moon (the moon goddess *Hina*), the great deity, the sun (the sun god *Tiki*) (and) the dark moon (the moon goddess *Hina Uri*), the month *Maru* (June for the major part), the sun, the darkness … appeared.'

2 (Br 3): **15-25 68 35 7 70 31 2 6 5-15 6 1 3** (= a reversed variant) **6 5 6 30 53 6 9 33** … *Ro(h)u Ono pe Tuu Pu: Make, Hina, a atua roa, a Tiki Hina uri, a atua, a ana Maru, a raa, a niva UA* …'The Pleiades (*Ono* = The Six [Stars]) and Aldebaran produced: (those were) the sun (the sun god *Makemake*), the moon (the moon goddess *Hina*), the great deity, the sun (the sun god *Tiki*) (and) the dark moon (the moon goddess *Hina Uri*), the brightness (the time near the winter solstice) of the month *Maru* (June for the major part), the sun, the darkness …'

A lunar eclipse before the day (the shine) of the winter solstice is described in these parallel records. Such a partial (almost total) eclipse was before the sunrise on June 20, A.D. 1796. Fedorova (1982: 50, 52) has read the combination of the glyphs **53** and **30** as the designation of the month *Maro*, but she has erroneously split the first sign into two parts representing hands (glyphs **15**). Glyph **53** *Maro* (*Maru*) depicts two extended arms, cf. Rapanui *maroa* 'fathom' and Maori *whakamārō* 'to extend, to stretch.' She has read glyphs **7 70** as *(h)etuu Pu* 'the star Aldebaran.'

In the first record this star is designated only with the word *tuu* 'star.' Hence, star-like signs in the local rock art can denote Aldebaran (see figure 3).

Consider a fragment on the Great Washington tablet (Sa 5), see figure 9.

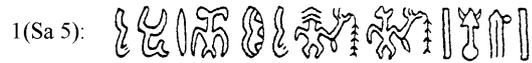

Figure 9

1 (Sa 5): **43 2 30-44 82 43 33/6-15 24 6-15 24 4 21 26-4** *Ma Hina Anakena, Pipiri; ma ua, Hora ari, Hora ari, atua ko Matua*. 'The moon of the month *Anakena* (July for the major part) (and) Canopus came; the rains, (then) the month *Hora-iti* (August for the major part) of the clear sun, the month *Hora-nui* (September for the major part) of the clear sun (and) the deity 'The Boat (β and α Centauri)' came.'

Old Rapanui *ma* 'to come' is comparable with Maori *ma* 'ditto.' Old Rapanui *ari* 'clear' correlates with Tahitian *ariari* 'ditto.'

For example, at first the moon rose (on July 22, A.D. 1845, 21:51), then Canopus rose (on July 23, A.D. 1845, 01:40). The heliacal rising of β Centauri occurred on September 2, A.D. 1845.

**On Another Rock Drawing**

Let us consider a panel at Tongariki (Lee 1992: 127, fig. 4.134) located in the eastern part of the island, see figure 10 (a segment of the drawing).

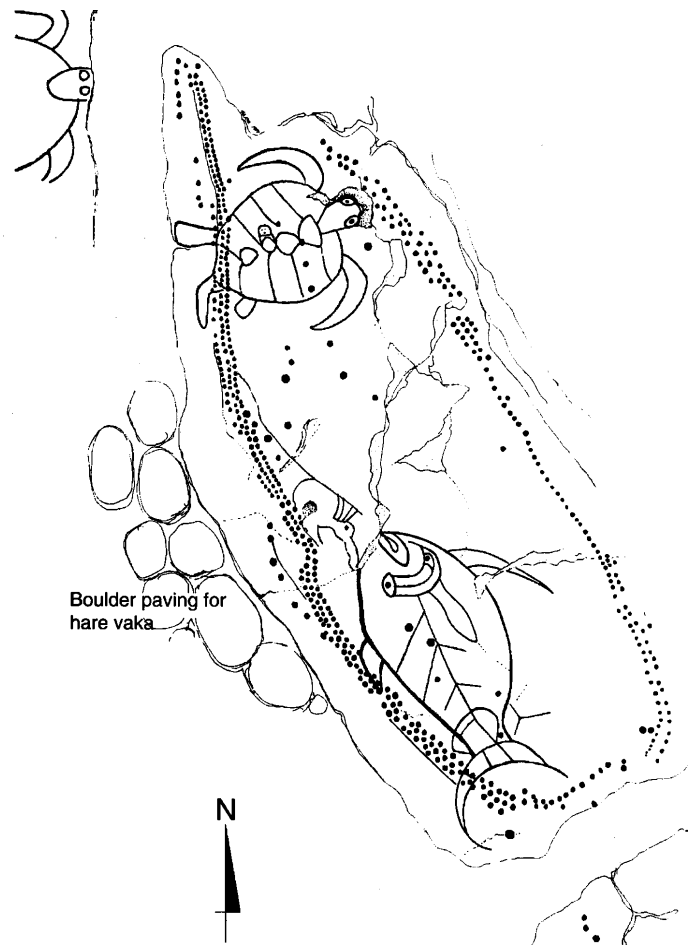

Figure 10.

Here a bird-man character (the image of one of elected bird-men) and many dots (calendar marks) are depicted. The number of the basic dots is around 428 (some short segments are seen badly). If the counting was conducted on the first night/day (*Hiro*) of the month *Maru* (*Maro*), the local New Year, we shall receive the date of the beginning of the month *Hora Iti* (August for the major part) of the second year. One can estimate that then a warrior of the eastern tribes won and became a sacral ruler or bird-man. The previous year could be the time of another bird-man from the western tribes. This result can be related to the statement of Métraux (1940: 340) that at the end of July the warriors gathered at Mataveri and Orongo and were involved in ceremonies of the bird cult.

## Conclusions

Different island cultures were closely related to each other in the distant past. However, it is apparent that the Maori (New Zealand)-Rapanui parallels are very significant. The personages *Makemake* and *Haua* were quite Polynesian deities. We could see that the bird-man rite was invented inside Polynesia. It is clear that this cult required the reliable astronomical observations of the sun, the moon, β and α Centauri, Canopus and Aldebaran.

# Appendix

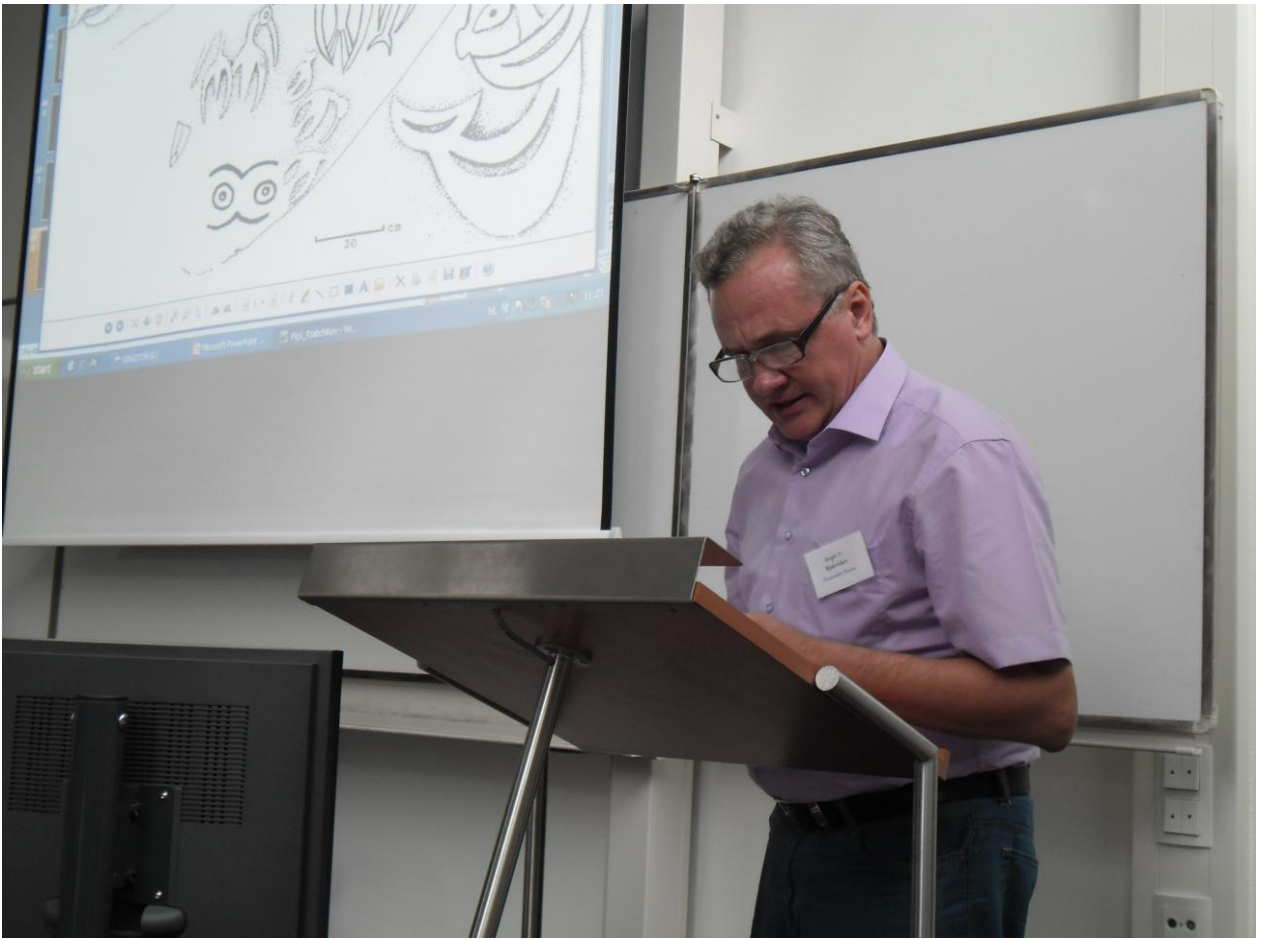

I was reading my paper.
I was warmly accepted by the
participants of the conference.